# Spectral window engineering for temporal leakage of synthetic gain waveform

Fuxin Guan[1,7,*], Nanyu Chen[1,7], Zemeng Lin[2,7], Wange Song[1,2], Shining Zhu[1], Tao Li[1,*], Shuang Zhang[2,3,4,5,6,*]

Email: guanfuxin@nju.edu.cn, taoli@nju.edu.cn, shuzhang@hku.hk

[1]National Laboratory of Solid State Microstructures, College of Engineering and Applied Sciences, Collaborative Innovation Center of Advanced Microstructures, Nanjing University, Nanjing, China

[2]New Cornerstone Science Laboratory, Department of Physics, University of Hong Kong, Hong Kong, China

[3]State Key Laboratory of Optical Quantum Materials, University of Hong Kong, Hong Kong, China

[4]Quantum Science Center of Guangdong-Hong Kong-Macao Great Bay Area, Shenzhen, China

[5]Department of Electrical & Electronic Engineering, University of Hong Kong, Hong Kong, China

[6]Materials Innovation Institute for Life Sciences and Energy (MILES), HKU-SIRI, Shenzhen, China

[7]These authors contributed equally: Fuxin Guan, Nanyu Chen, Zemeng Lin.

**Abstract:** Arbitrary waveform generation enables precise engineering of optical fields. As an alternative, the synthetic wave approach has been proposed to generate virtual-gain waveforms for loss compensation. However, the achievable synthetic gain is severely limited by the finite spectral bandwidth available in realistic experiments. This imposes a rectangular spectral window, causing abrupt spectral truncation and inducing severe temporal leakage in the time domain synthetic response. Here, we introduce spectral window engineering to fundamentally resolve this limitation. By employing smooth-edged spectral filters, such as the Hann and Blackman windows, temporal leakage is pushed deeply into the late-time regime, yielding a significantly extended effective lifetime and reduced residual damping. Using a coupled plasmonic resonator platform, we experimentally validate that our windowed approach suppresses residual damping by more than a factor of two compared to the standard rectangular baseline. Our findings establish a practical and robust spectral filtering strategy to realize high efficiency synthetic gain, with promising prospects for nanodetection and advanced spectroscopy.

**Introduction:**

Arbitrary waveform generation enables flexible tailoring of optical temporal profiles and phase evolution, underpinning a broad range of applications from high-speed optical communications to ultrafast spectroscopic detection [1–15]. As a coherent broadband light source consisting of uniformly spaced, precisely tunable frequency lines, the optical frequency comb provides a high accuracy platform for synthesizing arbitrary optical waveforms [16–26]. Complex frequency excitations, a class of specialized temporal waveforms featuring exponentially decaying or growing profiles, have emerged as a distinctive tool to introduce effective gain or loss in the time domain without modifying the intrinsic material properties [27–56], enabling a wealth of exotic phenomena inaccessible under conventional real frequency excitations [57–64].

Recently, the synthetic wave approach has emerged as an alternative scheme to construct arbitrary temporal waveforms [65–68]. Instead of relying on direct excitation, it enables flexibly temporal waveform synthesis using solely multiple real frequency components, and has been widely adopted to synthesize temporal decays of complex frequency excitations for loss compensation, facilitating a variety of loss compensated functionalities in nanodetection systems [69–74]. Compared with conventional waveform generation techniques, this approach holds great promise for concise ultrafast waveform synthesis.

Nevertheless, its performance is limited by two fundamental factors. First, synthetic complex frequency waves inherently exhibit an abrupt temporal onset, which unavoidably excites intrinsic resonance of the system as spurious noise into the temporal signal. This specific challenge has been resolved by employing higher-order synthetic gain [66], which effectively smooths the temporal onset of the excitation. Second, synthesizing a complex frequency signal requires a broad spectral range, especially in high-loss systems that demand a rapidly decaying temporal envelope. However, the available spectral bandwidth is strictly bounded by experimental setup. This limitation imposes an effective rectangular spectral cutoff, which produces a nonlocal temporal kernel with a 1/t asymptotic decay. The resulting severe temporal leakage quickly degrades the fast decaying synthetic signal, constituting a critical bottleneck for synthetic gain implementations in high-loss systems that cannot be addressed merely by higher-order synthetic gain.

Here, we tackle this spectral cutoff problem using spectral window filtering and experimentally validate the enhancement using a double resonance plasmonic platform. Theoretically, the residual damping following synthetic gain exhibits an inverse temporal dependence governed by the scaling law $\gamma_{eff}(N,t) \approx (N+2)/t$, where $N$ denotes the synthetic gain order. Ideally, this loss decreases monotonically over time. In practice, however, the hard rectangular cutoff induces temporal leakage that overwhelms the signal at long times, breaking this scaling relation.

Spectral window filtering, such as Hann and Blackman windows, fundamentally addresses this by smoothing the spectral truncation [75–77], which accelerates the decay profile of temporal leakage from $1/t$ to $1/t^3$. By systematically comparing fix spectral window profiles, we demonstrate that smoother-edged windows yield a substantially longer synthetic gain lifetime. Specifically, Hann, Blackman, and Tukey windows successfully restore the inverse time-scaling law in the long-time regime. Our experimental measurements confirm this breakthrough, showing that proper spectral windowing reduces residual damping by more than half compared to the rectangular baseline, thereby delivering a massive boost in loss compensation efficiency.

**Results:**

We begin with the frequency domain representation of the synthetic complex frequency wave, given by

$$E_{syn}(t) = \int_{-\infty}^{\infty} \frac{W(\omega)e^{-i\omega t}}{2\pi i(\tilde{\omega}_0-\omega)} d\omega \tag{1}$$

where $W(\omega) = \begin{cases} 1 & where\ |\omega-\tilde{\omega}_0| < \omega_\Delta \\ 0 & else \end{cases}$ indicates the rectangular spectral window function applied to the finite synthesis bandwidth with value of $2\omega_\Delta$, and $\tilde{\omega}_0 = \omega_0 - i\beta$ denotes the complex frequency. By the convolution theorem, the product of the window and the Lorentzian kernel maps to a temporal convolution in the time domain,

$$E_{syn}(t) = \int_0^{\infty} \tilde{W}(t-\tau)e^{-i\tilde{\omega}_0\tau} d\tau \tag{2}$$

where $\widetilde{W}(t)$ is the temporal kernel obtained from the Fourier transform of $W(\omega)$, and the lower integration bound at $\tau = 0$ originates from the step onset of the truncated complex frequency waveform. This convolution indicates that any extended tail in $\widetilde{W}(t-\tau)$ couples strong early time components into the late time response, producing significant spurious leakage at large time.

We here investigate six distinct spectral window functions, as shown in Fig. 1(a). The rectangular window features an abrupt cutoff with a flat unity response inside the available frequency range. The Tukey window combines a flat central region with cosine-tapered edges $W_{Tukey}(\omega) = \begin{cases} 1 & |\omega-\omega_0| \le 0.5\omega_\Delta \\ 0.5\left[1+cos\left(\frac{\pi(|\omega-\omega_0|-0.5\omega_\Delta)}{0.5\omega_\Delta}\right)\right] & 0.5\omega_\Delta \le |\omega-\omega_0| \le \omega_\Delta \end{cases}$, the Hann window $W_{Hann}(\omega) = 0.5 + 0.5cos\left(\frac{\pi(\omega-\omega_0)}{\omega_\Delta}\right)$, Hamming window $W_{Hamming}(\omega) = 0.54 + 0.46cos\left(\frac{\pi(\omega-\omega_0)}{\omega_\Delta}\right)$ and Blackman window $W_{Blackman}(\omega) = 0.42 + 0.5cos\left(\frac{\pi(\omega-\omega_0)}{\omega_\Delta}\right) + 0.08cos\left(\frac{2\pi(\omega-\omega_0)}{\omega_\Delta}\right)$ represent cosine series designs, with all available frequency bandwidths are located within $|\omega-\omega_0| < \omega_\Delta$. Finally, the Gaussian window $W_{Gaussian}(\omega) = \exp\left(-\frac{9(\omega-\omega_0)^2}{2\omega_\Delta^2}\right)$ adopts a Gaussian envelope truncated at the bandwidth boundaries.

The corresponding temporal kernels of all windows are compared in Fig. 1(b) on a logarithmic amplitude scale. The rectangular window yields a sinc kernel $\widetilde{W}_{rect}(t) = 2\sin(\omega_\Delta t)e^{-i\omega_0 t}/t$ with prominent oscillatory sidelobes that decay as $1/t$, producing the slowest decaying among all tested windows. The Hamming window slightly accelerate the decay, but its nonzero amplitude at the spectral band edges introduces a residual discontinuity that slows the asymptotic decay. The truncated Gaussian window exhibits similarly slow decay due to the amplitude discontinuity. The Tukey window, with smoothed edges, reduces the sidelobe below that of the Gaussian window. Crucially, the Hann window $\widetilde{W}(t) = \frac{\sin(\omega_\Delta t)e^{-i\omega_0 t}}{t(1-\omega_\Delta^2 t^2/\pi^2)}$, featuring zero endpoints and continuous first derivatives at the band edges, follows a steeper $1/t^3$ asymptotic decay, leading to a significantly lower kernel amplitude in the long-time regime.

Simultaneously, the Blackman window achieves the same $1/t^3$ algebraic decay but maintains an even smoother second derivative at boundaries compared to Hann window, ultimately yielding the lowest long-time sidelobes across the entire tested set.

Fig. 1(c) shows the normalized spectra of the windowed synthetic complex-frequency waves. The rectangular window spectrum retains sharp discontinuities at the band edges inherited from its hard spectral cutoff. In contrast, all other windows effectively soften the spectral truncation, eliminating the abrupt edge jumps responsible for generating fast-decaying temporal leakage.

The resulting temporal evolution of the synthetic complex frequency waveforms is plotted in Fig. 1(d), with the ideal exponential decay included as a dashed reference. The rectangular window causes the earliest deviation from the ideal decay trajectory and settles at the highest temporal leakage. Consistent with the temporal kernel trends in Fig. 1(b), windows with faster-decaying kernels consistently yield lower long time signal amplitudes and extend the duration of faithful exponential decay. The Blackman window achieves the lowest level of temporal leakage at long times, followed by the Hann, Tukey, Gaussian and Hamming windows, with the rectangular window performing the worst. This trend confirms that spectral edge smoothness is the governing factor for temporal leakage.

We next investigate temporal leakage in synthetic response rather than merely the waveform. In an electromagnetic system, linear response to an external electromagnetic field is described by a retarded response function. The time-domain polarization $P(t)$ induced by an electric field $E_{in}(t)$ is given by $P(t) = \int_{-\infty}^{t} \chi(t-\tau) E_{in}(\tau) d\tau$, where the susceptibility $\chi(t) = \chi_0 e^{-i\Omega t - \gamma t}$ accounts for the resonance frequency $\Omega$ and damping rate $\gamma$. Here, we only consider the positive frequency resonance for convenience. The upper limit $t$ in the time integral enforces causality, indicating that $P(t)$ depends only on the field at earlier times. Hence it can be written as, $P(t) = \int_{-\infty}^{t} \chi_0 e^{-(i\Omega+\gamma)(t-\tau)} E_{in}(\tau) d\tau$, with $t > \tau$. If the input signal is harmonic, $E_{in}(\tau) \propto e^{-i\omega_0 \tau}$, the resulting polarization takes the form, $P(t) \propto \chi_0 e^{-i\omega_0 t}/(\omega_0 - \Omega + i\gamma)$. Thus, the susceptibility can be derived, $\chi(\omega_0) \propto (\omega_0 - \Omega + i\gamma)^{-1}$, which exhibits the familiar Lorentzian lineshape.

To investigate the synthetic loss compensation, we employ an $N^{\text{th}}$-order synthetic gain envelope for the input field [66,72]:

$$E_{in}(N,t) = \theta(t)t^N e^{-(i\omega_0+\beta)t} \tag{3}$$

where $\beta$ represents the virtual gain, $N \geq 0$ denotes the order of synthetic gain, and $\theta(t)$ is the Heaviside step function. Substituting Eq. (3) into the polarization equation, we derive the analytical expression of the effective susceptibility $\chi_{eff}(N,t)$, with the corresponding results provided in the Supplementary Information. For simplicity, we select the virtual gain to match the value of natural damping ($\beta = \gamma$) and obtain $P(N,t) = \chi_0 e^{-(i\Omega+\gamma)t} \int_0^t \tau^N e^{-s\tau} d\tau$, with $\Delta\omega_0 = \omega_0 - \Omega$, and $s = i\Delta\omega_0$. Evaluating the integral yields,

$$P(N,t) = \chi_0 e^{-(i\Omega+\gamma)t} s^{-N-1} \Gamma(N+1,st) \tag{4}$$

where $\Gamma(N+1,st) = \int_0^{st} \epsilon^N e^{-\epsilon} d\epsilon$ is the incomplete gamma function. The effective susceptibility is therefore, $\chi_{eff}(N,t) = P(N,t)/E_{in}(N,t) = \chi_0 e^{st} s^{-N-1} t^{-N} \Gamma(N+1,st)$. Expanding $\chi_{eff}(N,t)$ for small $st$, one obtains the series $\chi_{eff}(N,t) = \chi_0 \frac{t}{N+1}\left[1 + \frac{st}{N+2} + \cdots\right]$. Near resonance ($|\gamma_{eff}| > |s|$), we employ Lorentzian-like form to expand effective susceptibility [72], $\chi_{eff} \propto 1/(\gamma_{eff} - s)$. Expanding this in $s$ gives, $\chi_{eff} \propto \frac{1}{\gamma_{eff}}(1 + \frac{s}{\gamma_{eff}} + \cdots)$. Matching the two series of $\chi_{eff}$ (e.g., via the Padé approximation) leads to the following estimate for the dynamics of the residual damping,

$$\gamma_{eff}(N,t) \approx (N+2)/t \tag{5}$$

It shows that the residual damping scales inversely with time. Longer time acquires better loss compensation efficiency.

Here we investigate this scaling law in a Lorentzian model using synthetic wave approach, with $\Omega = 2$ and $\gamma = 0.025$. By introducing a rectangular window function $W(\omega)$, the synthetic polarization can be derived as,

$$P_{syn}(N,\tilde{\omega}_0,t) = N! \int_{-\infty}^{\infty} W(\omega)\chi(\omega)e^{-i\omega t}/[2\pi(i\tilde{\omega}_0 - i\omega)^{N+1}]d\omega \tag{6}$$

where $\chi(\omega) = -\omega_p(\omega - \Omega + i\gamma)^{-1}$, $\widetilde{\omega}_0 = \omega_0 - i\gamma$, and $t > 0$. Accordingly, the synthetic susceptibility can be expressed as $\chi_{syn}(N, \widetilde{\omega}_0, t) \equiv t^{-N} e^{i\widetilde{\omega}_0 t} P_{syn}(N, \widetilde{\omega}_0, t)$. Using Eq. (6), we plot the synthetic polarization for $N = 1$ and $N = 2$ at a fixed frequency cutoff at $\omega_\Delta = 1$, with the results are depicted in Figs. 2(a, e). It is observed that the higher-order result exhibits a cleaner early time signals and a delayed onset of temporal leakage, consistent with the findings established in Ref. [66].

To mitigate temporal leakage at large times, we replace the rectangular window with a Hann window, $W(\omega) = \begin{cases} 1 + \cos(\pi(\omega - \Omega)/\omega_\Delta) & where\ |\omega - \Omega| < \omega_\Delta \\ 0 & else \end{cases}$. Substituting this Hann function into Eq. (6), we obtain the temporal evolutions of $Im[\chi_{syn}]$ for $N = 1$ and $N = 2$, shown in Figs. 2(b) and 2(f). The difference fields, obtained by subtracting the rectangular window results from the corresponding Hann window results, are presented in Figs. 2(c, g), respectively. It is evident that the Hann window exclusively suppresses temporal leakage and prolongs the lifetime of synthetic susceptibility, without altering its intrinsic short time profile.

In Fig. 2(d), we extract inverse residual damping of $|\chi_{syn}(\omega, t)|^2$ in Figs. 2(a) and 2(b), denoted by the blue dashed line (rectangular window) and red line (Hann window), respectively. Data points for the blue dashed line at elapsed times exceeding 600 are omitted because of the pronounced temporal leakage. The corresponding inverse residual damping derived from Figs. 2(e) and 2(f) as presented in Fig. 2(h). A comparison of the paired red and blue curves in Figs. 2(d) and 2(h) reveals that the Hann window significantly prolongs the lifetime of the effective susceptibilities for both orders. Furthermore, the residual damping is reduced by at least half compared to the rectangular window baseline. This demonstrates that spectral window filtering provides an order independent improvement to the loss compensation efficiency of synthetic gain. Quantitively, under the Hann window, inverse residual damping converges to an identical value of 333 for both orders, whereas the uncompensated value equals to 20. Consequently, the residual damping can be minimized to $\gamma_{eff} = 6\%\gamma = 0.15\%\omega_\Delta$ under such condition.

We use the measured transmission spectrum of a plasmon-induced transparency (PIT) system to illustrate the temporal evolution of the effective damping and to assess the improvement enabled by the spectral windows. The fabricated sample, shown in Fig. 3(a), supports two coupled resonant modes [78]. Owing to strong plasmonic loss, the corresponding transmission dips are substantially broadened. The measured transmission spectrum $T_M$ satisfies the Kramers–Kronig relations, so its amplitude and phase $\varphi_M$ are causally linked. The complex response $t_M(\omega) = e^{i\varphi_M}\sqrt{T_M}$ can be well explained by the superposition of a smooth background and two Lorentzian resonances with distinct frequencies [79].

By removing the background contribution, we isolate the polarization response $P(\omega)$, as shown by the black line in Fig. 3(b), and construct the synthetic polarization $P_{syn}(\tilde{\omega}, t)$ for $N$=2 via Eq. (6). Using a rectangular spectral window (Fig. 3c), the temporal evolution initially reproduces the expected separation of the two resonances. However, as time progresses, temporal leakage rapidly develops and eventually dominates the response. Replacing the rectangular window with a Hann, Blackman or Tukey window filter qualitatively alter this picture. As shown in Figs. 3(d)-3(f), the temporal evolution becomes markedly cleaner, and the resonant features remain well-defined over extended durations. We extract the polarization results at the positions of the white dashed lines from Figs. 3(c)-3(f) and plot them in Fig. 3(b). All windowed spectra exhibit significantly higher quality factors than the direct measurement. They feature sharper resonant lineshapes as a result of more effective loss compensation, with the three windowed cases demonstrating largely comparable performance.

To quantify this effect, we extract the residual damping from the right peak of polarization intensities. We compare the inverse residual damping for all four windowed cases, as plotted in Fig. 3(g). The Hann-, Blackman-, and Tukey-filtered data all increasing monotonically to $\gamma_{eff} = 33cm^{-1}$, corresponding to $0.2\beta$. In contrast, the rectangular window data (dark blue) rapidly deviate from this monotonically increasing behavior, exhibiting a minimum damping of $0.5\beta$.

**Conclusion**

In summary, we have identified and resolved the temporal leakage limitation inherent in the synthetic wave approach. We analytically derived an inverse-time scaling law, $\gamma_{eff}(N,t) = (N+2)/t$, which describes the residual damping after applying synthetic gain. However, under realistic finite-bandwidth conditions, this ideal scaling is severely disrupted in the long-time regime by temporal leakage induced by rectangular spectral truncation. To overcome this fundamental bottleneck, we introduced a spectral window engineering technique. By employing smooth-edged filters, such as the Hann, Blackman, and Tukey windows, we accelerate the decay of the temporal kernel, successfully suppressing temporal leakage and restoring the ideal scaling law at late times. Our experimental measurements validate that proper spectral windowing suppresses residual damping by more than a factor of two compared to the standard rectangular baseline. Crucially, this performance enhancement is order-independent, acting as a purely additive upgrade to any synthetic gain order. This robust spectral window engineering framework paves the way for transformative breakthroughs in arbitrary waveform generation, advanced spectroscopy, and optical nanodetection.

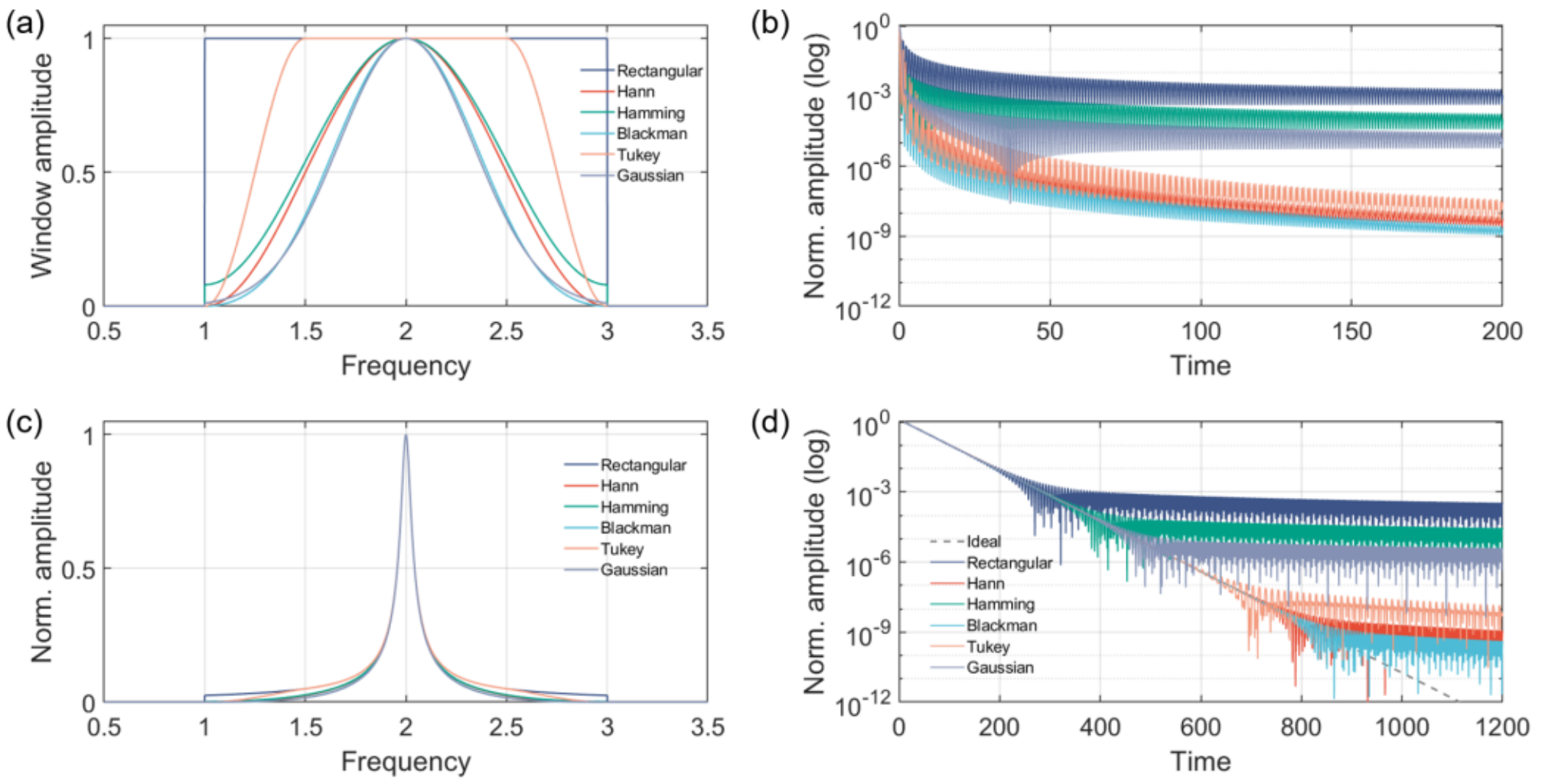

**Figure 1**. Performance comparison of six spectral window functions. (a) Spectral profile of the rectangular, Hann, Hamming, Blackman, Tukey and Gaussian windows. (b) Corresponding six temporal kernels on a logarithmic amplitude scale. (c) Normalized spectra of windowed synthetic complex frequency waveforms. (d) Temporal evolution of the synthesized waveforms on a logarithmic scale, with the ideal exponential decay included as a dashed reference.

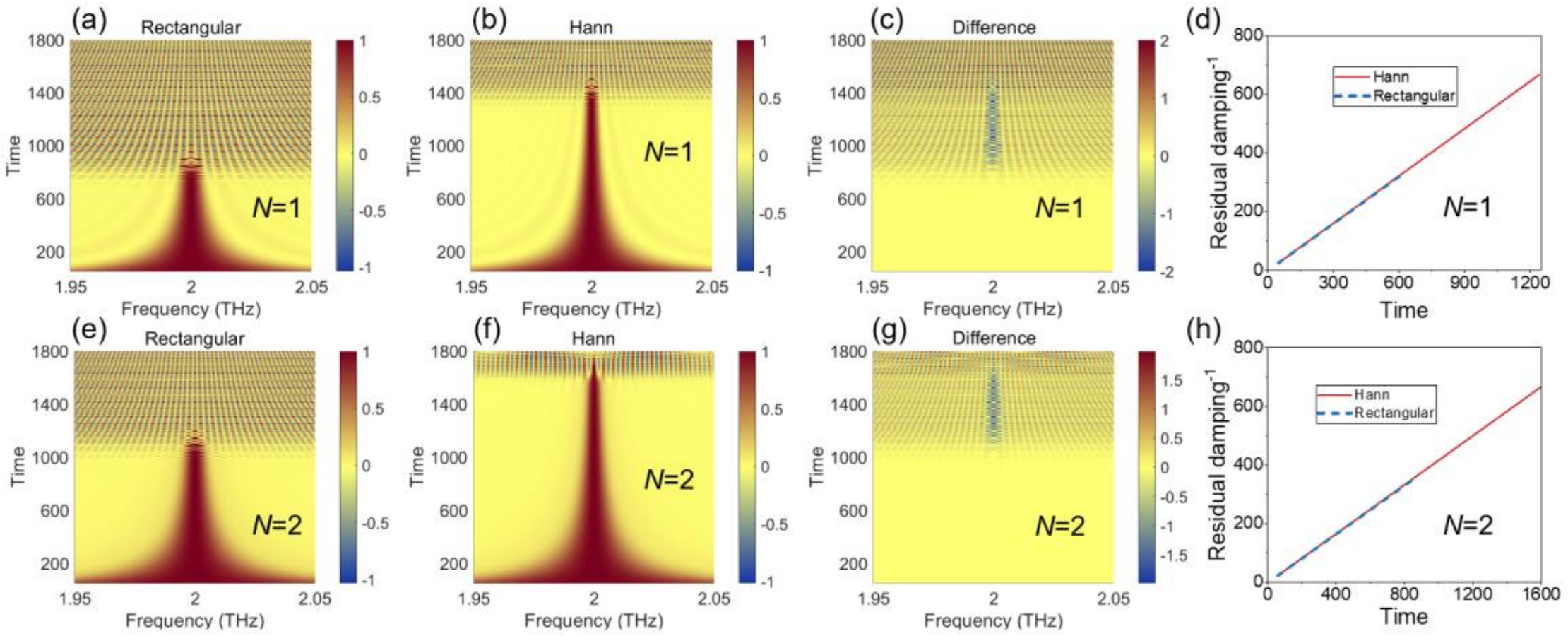

**Figure 2**. Temporal evolutions of imaginary synthetic susceptibility under rectangular and Hann windows. (a) and (b) Frequency and time maps of $Im[\chi_{syn}(N,\omega,t)]$ for *N*=1 under rectangular and Hann window functions, respectively. (c) Difference field derived from (a) minus (b). (d) Residual damping extracted from $|\chi_{syn}(N,\omega,t)|^2$ of (a) and (b) versus elapsed time. (e)-(h) present the corresponding results for *N*=2.

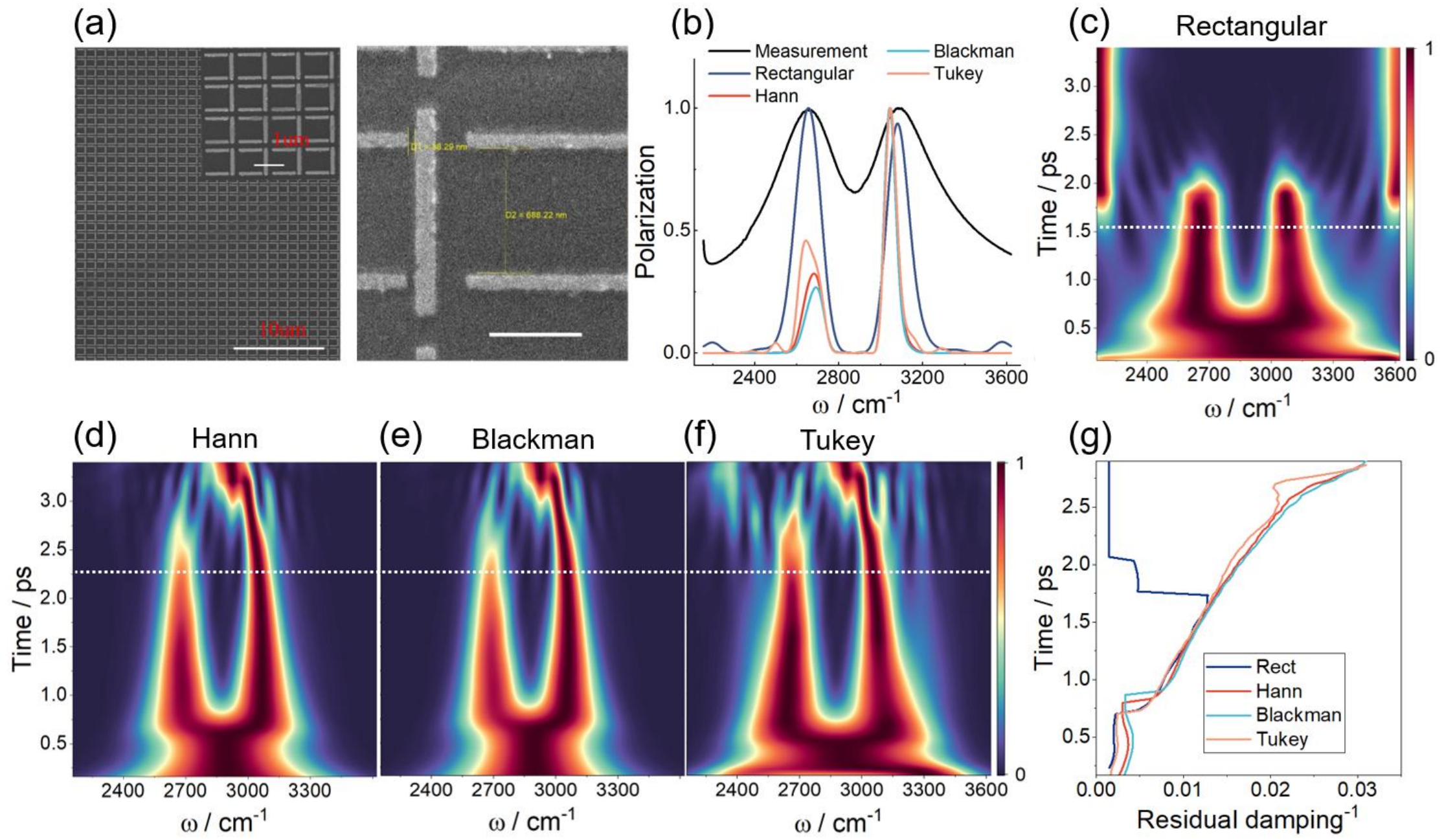

**Figure 3**. Measured transmission spectrum of the plasmon-induced transparency (PIT) metamaterial, and synthesized polarizations obtained using various window filters. (a) Scanning electron microscopy image of the fabricated sample, including magnified views of unit cells. (b) Polarization intensity $|P(\omega)|^2$ extracted from the measured transmission spectrum (black line). The synthesized polarizations $|P_{syn}(\widetilde{\omega},t)|^2$ obtained with rectangular, Hann, Blackman and Tukey windows are plotted as colored lines, the corresponding temporal snapshots are marked by the white dashed lines in (c-f). (c-f) Temporal evolutions of $|P_{syn}(\widetilde{\omega},t)|^2$ obtained using (c) a rectangular window, (d) a Hann window, (e) a Blackman window, and (f) a Tukey window, with the virtual gain set to $\beta = 160cm^{-1}$. (g) Residual damping of right resonance peaks in (c-f), evaluated from the FWHM, plotted as color-matched curves.

***Acknowledgments***

The authors acknowledge the financial support from National Natural Science Foundation of China (Grant Nos. 62288101, 62325504), the New Cornerstone Science Foundation, the Research Grants Council of Hong Kong (AoE/P-502/20, STG3/E-704/23-N, 17309021), the National Key Research and Development Program of China (No. 2022YFA1404301), Guangdong Provincial Quantum Science Strategic Initiative (GDZX2204004, GDZX2304001). F. G. thanks the support from Dengfeng Project B of Nanjing University.

The authors declare no conflicts of interest.

***Data availability***

The data are available from the authors upon reasonable request.

**References:**

[1] A. M. Weiner, D. E. Leaird, G. P. Wiederrecht, and K. A. Nelson, Femtosecond Pulse Sequences Used for Optical Manipulation of Molecular Motion, Science (1979). **247**, 1317 (1990).

[2] R. Holzwarth, T. Udem, T. W. Hänsch, J. C. Knight, W. J. Wadsworth, and P. J. St Russell, Optical Frequency Synthesizer for Precision Spectroscopy, Phys. Rev. Lett. **85**, 2264 (2000).

[3] A. M. Weiner, Femtosecond pulse shaping using spatial light modulators, Review of Scientific Instruments **71**, 1929 (2000).

[4] N. Dudovich, D. Oron, and Y. Silberberg, Single-pulse coherently controlled nonlinear Raman spectroscopy and microscopy, Nature **418**, 512 (2002).

[5] P. Michel, W. Benz, and D. C. Richardson, Attosecond control of electronic processes by intense light fields, Nature **421**, 611 (2003).

[6] D. Goswami, Optical pulse shaping approaches to coherent control, Phys. Rep. **374**, 385 (2003).

[7] Z. Jiang, C. Bin Huang, D. E. Leaird, and A. M. Weiner, Optical arbitrary waveform processing of more than 100 spectral comb lines, Nat. Photonics **1**, 463 (2007).

[8] N. K. Fontaine, R. P. Scott, J. Cao, A. Karalar, W. Jiang, K. Okamoto, J. P. Heritage, B. H. Kolner, and S. J. B. Yoo, 32 PhaseÃ 32 Amplitude Optical Arbitrary Waveform Generation, 2007.

[9] Z. Jiang, D. S. Seo, D. E. Leaird, and A. M. Weiner, Spectral Line-by-Line Pulse Shaping, 2005.

[10] B. J. Bloom, T. L. Nicholson, J. R. Williams, S. L. Campbell, M. Bishof, X. Zhang, W. Zhang, S. L. Bromley, and J. Ye, An optical lattice clock with accuracy and stability at the 10-18 level, Nature **506**, 71 (2014).

[11] M. H. Khan, H. Shen, Y. Xuan, L. Zhao, S. Xiao, D. E. Leaird, A. M. Weiner, and M. Qi, Ultrabroad-bandwidth arbitrary radiofrequency waveform generation with a silicon photonic chip-based spectral shaper, Nat. Photonics **4**, 117 (2010).

[12] M. Veli, D. Mengu, N. T. Yardimci, Y. Luo, J. Li, Y. Rivenson, M. Jarrahi, and A. Ozcan, Terahertz pulse shaping using diffractive surfaces, Nature Communications **12**, (2021).

[13] Y. Schrödel et al., Acousto-optic modulation of gigawatt-scale laser pulses in ambient air, Nat. Photonics **18**, 54 (2024).

[14] D. Drayss et al., Optical arbitrary waveform generation (OAWG) using actively phase-stabilized spectral stitching, Light Sci. Appl. **14**, (2025).

[15] Y. Guan, G. Wang, Y. Zhi, J. Chen, L. Li, J. Zhang, and J. Yao, Temporal point-by-point arbitrary waveform synthesis beyond tera sample per second, Nature Communications **16**, (2025).

[16] D. J. Jones, S. A. Diddams, J. K. Ranka, A. Stentz, R. S. Windeler, J. L. Hall, and S. T. Cundiff, Carrier-Envelope Phase Control of Femtosecond Mode-Locked Lasers and Direct Optical Frequency Synthesis, Science (1979). **288**, 635 (2000).

[17] S. A. Diddams, D. J. Jones, J. Ye, S. T. Cundiff, J. L. Hall, J. K. Ranka, R. S. Windeler, R. Holzwarth, T. Udem, and T. W. Hänsch, Direct Link between Microwave and Optical Frequencies with a 300 THz Femtosecond Laser Comb, Phys. Rev. Lett.

**84**, 5102 (2000).
[18] P. Del'Haye, A. Schliesser, O. Arcizet, T. Wilken, R. Holzwarth, and T. J. Kippenberg, Optical frequency comb generation from a monolithic microresonator, Nature **450**, 1214 (2007).
[19] S. T. Cundiff and J. Ye, Colloquium: Femtosecond Optical Frequency Combs, 2003.
[20] F. Ferdous, H. Miao, D. E. Leaird, K. Srinivasan, J. Wang, L. Chen, L. T. Varghese, and A. M. Weiner, Spectral line-by-line pulse shaping of on-chip microresonator frequency combs, Nat. Photonics **5**, 770 (2011).
[21] S. Tammaro, O. Pirali, P. Roy, J. F. Lampin, G. Ducournau, A. Cuisset, F. Hindle, and G. Mouret, High density terahertz frequency comb produced by coherent synchrotron radiation, Nature Communications **6**, (2015).
[22] H. Guo, C. Herkommer, A. Billat, D. Grassani, C. Zhang, M. H. P. Pfeiffer, W. Weng, C. S. Brès, and T. J. Kippenberg, Mid-infrared frequency comb via coherent dispersive wave generation in silicon nitride nanophotonic waveguides, Nat. Photonics **12**, 330 (2018).
[23] A. L. Gaeta, M. Lipson, and T. J. Kippenberg, Photonic-chip-based frequency combs, Nat. Photonics **13**, 158 (2019).
[24] N. Picqué and T. W. Hänsch, Frequency comb spectroscopy, Nat. Photonics **13**, 146 (2019).
[25] A. Roy, L. Ledezma, L. Costa, R. Gray, R. Sekine, Q. Guo, M. Liu, R. M. Briggs, and A. Marandi, Visible-to-mid-IR tunable frequency comb in nanophotonics, Nature Communications **14**, (2023).
[26] M. Liu, R. M. Gray, L. Costa, C. R. Markus, A. Roy, and A. Marandi, Mid-infrared cross-comb spectroscopy, Nature Communications **14**, (2023).
[27] A. Archambault, T. V. Teperik, F. Marquier, and J. J. Greffet, Surface plasmon Fourier optics, Phys. Rev. B Condens. Matter Mater. Phys. **79**, (2009).
[28] S. Kim, A. Krasnok, and A. Alù, Complex-frequency excitations in photonics and wave physics, Science (1979). **387**, eado4128 (2025).
[29] A. Archambault, M. Besbes, and J. J. Greffet, Superlens in the time domain, Phys. Rev. Lett. **109**, 097405 (2012).
[30] H. Li, A. Mekawy, A. Krasnok, and A. Alù, Virtual Parity-Time Symmetry, Phys. Rev. Lett. **124**, 193901 (2020).
[31] Q. Zhao, M. Li, Z. Chen, S. Kim, H. Li, J. Xu, and A. Alù, Virtual Gain for Coherent Perfect Absorption in Ultrathin Excitonic Slabs, ACS Photonics **12**, 5399 (2025).
[32] Y. Ra'Di, A. Krasnok, and A. Alù, Virtual Critical Coupling, ACS Photonics **7**, 1468 (2020).
[33] S. Lepeshov and A. Krasnok, Virtual optical pulling force, Optica **7**, 1024 (2020).
[34] D. N. Basov and M. M. Fogler, *"The Unreasonable Effectiveness of Mathematics" in Evading Polaritonic Losses*, Nature Materials.
[35] Q. Cheng and T. Li, Complex-frequency waves: beat loss and win sensitivity, Light Sci. Appl. **13**, 40 (2024).
[36] Richard M. More and Edward Gerjuoy, Properties of Resonance Wave Functions, Phys. Rev. A **7**, 1973 (1973).
[37] E. S. C. Ching, P. T. Leung, A. Maassen Van Den Brink, W. M. Suen, S. S. Tong, and

K. Young, Quasinormal-mode expansion for waves in open systems, Rev. Mod. Phys. **70**, 1545 (1998).
[38] P. Lalanne, W. Yan, K. Vynck, C. Sauvan, and J. P. Hugonin, Light Interaction with Photonic and Plasmonic Resonances, Laser Photon. Rev. **12**, 1700113 (2018).
[39] W. Yan, R. Faggiani, and P. Lalanne, Rigorous modal analysis of plasmonic nanoresonators, Phys. Rev. B **97**, 205422 (2018).
[40] J. Yang, H. Giessen, and P. Lalanne, Simple analytical expression for the peak-frequency shifts of plasmonic resonances for sensing, Nano Lett. **15**, 3439 (2015).
[41] C. Sauvan, J. P. Hugonin, I. S. Maksymov, and P. Lalanne, Theory of the spontaneous optical emission of nanosize photonic and plasmon resonators, Phys. Rev. Lett. **110**, 237401 (2013).
[42] P. T. Leung, S. Y. I.iu, and K. Young, Completeness and orthogonahty of qnaslnormal modes xn leaky optical cavities, Phys. Rev. A (Coll. Park). **49**, 3057 (1994).
[43] C. Tao, Y. Zhong, and H. Liu, Quasinormal Mode Expansion Theory for Mesoscale Plasmonic Nanoresonators: An Analytical Treatment of Nonclassical Electromagnetic Boundary Condition, Phys. Rev. Lett. **129**, (2022).
[44] C. Tao, J. Zhu, Y. Zhong, and H. Liu, Coupling theory of quasinormal modes for lossy and dispersive plasmonic nanoresonators, Phys. Rev. B **102**, (2020).
[45] P. Lalanne et al., Quasinormal mode solvers for resonators with dispersive materials, Journal of the Optical Society of America A **36**, 686 (2019).
[46] H. S. Tetikol and M. I. Aksun, Enhancement of Resolution and Propagation Length by Sources with Temporal Decay in Plasmonic Devices, Plasmonics **15**, 2137 (2020).
[47] Y. D. Chong, L. Ge, H. Cao, and A. D. Stone, Coherent perfect absorbers: Time-reversed lasers, Phys. Rev. Lett. **105**, 053901 (2010).
[48] D. G. Baranov, A. Krasnok, and A. Alù, Coherent virtual absorption based on complex zero excitation for ideal light capturing, Optica **4**, 1457 (2017).
[49] T. Delage, J. Sokoloff, O. Pascal, V. Mazières, A. Krasnok, and T. Callegari, Plasma Ignition via High-Power Virtual Perfect Absorption, ACS Photonics **10**, 3781 (2023).
[50] K. L. Tsakmakidis, T. W. Pickering, J. M. Hamm, A. F. Page, and O. Hess, Completely stopped and dispersionless light in plasmonic waveguides, Phys. Rev. Lett. **112**, 167401 (2014).
[51] S. Kim, S. Lepeshov, A. Krasnok, and A. Alù, Beyond Bounds on Light Scattering with Complex Frequency Excitations, Phys. Rev. Lett. **129**, (2022).
[52] D. Trivedi, A. Madanayake, and A. Krasnok, Revealing invisible scattering poles with complex-frequency signals, J. Appl. Phys. **137**, 243103 (2025).
[53] C. Rasmussen, M. I. N. Rosa, J. Lewton, and M. Ruzzene, A Lossless Sink Based on Complex Frequency Excitations, Advanced Science **10**, 2301811 (2023).
[54] J. Huang, J. Hu, and Z. Yang, Complex frequency detection in a subsystem, Commun. Phys. **9**, 84 (2026).
[55] M. Wu, M. Weng, Z. Chi, S. Zheng, F. Liu, C. Shang, B. Zhang, Q. Zhao, Y. Meng,

and J. Zhou, Gain-and-Loss-Free Metamaterials Exhibiting Non-Hermitian Non-Bloch Effects, Phys. Rev. Lett. **136**, 106603 (2026).

[56] Juntao Huang, Kun Ding, Jiangping Hu, and Zhesen Yang, Complex frequency fingerprint: Experimentally accessible method for detecting complex-valued eigenfrequencies, Phys. Rev. B **113**, 075128 (2026).

[57] S. Zhang, C. Liu, S. Zhou, C. S. Chuu, M. M. T. Loy, and S. Du, Coherent control of single-photon absorption and reemission in a two-level atomic ensemble, Phys. Rev. Lett. **109**, (2012).

[58] M. Bader, S. Heugel, A. L. Chekhov, M. Sondermann, and G. Leuchs, Efficient coupling to an optical resonator by exploiting time-reversal symmetry, New J. Phys. **15**, (2013).

[59] J. Wenner et al., Catching time-reversed microwave coherent state photons with 99.4% absorption efficiency, Phys. Rev. Lett. **112**, (2014).

[60] J. Hinney, S. Kim, G. J. K. Flatt, I. Datta, A. Alù, and M. Lipson, Efficient excitation and control of integrated photonic circuits with virtual critical coupling, Nat. Commun. **15**, 1 (2024).

[61] B. Xue, R. Zhang, Y. Zhu, Y. Sun, X. Chen, A. Alù, and W. Wan, Lasing-like dynamics with virtual gain driven by complex-frequency excitations, Nat. Commun. (2026).

[62] G. Trainiti, Y. Ra'di, M. Ruzzene, and A. Alù, Coherent virtual absorption of elastodynamic waves, Sci. Adv. **5**, eaaw3255 (2019).

[63] Z. Gu, H. Gao, H. Xue, J. Li, Z. Su, and J. Zhu, Transient non-Hermitian skin effect, Nat. Commun. **13**, (2022).

[64] S. Kim, Y. G. Peng, S. Yves, and A. Alù, Loss Compensation and Superresolution in Metamaterials with Excitations at Complex Frequencies, Phys. Rev. X **13**, 041024 (2023).

[65] F. Guan, X. Guo, K. Zeng, S. Zhang, Z. Nie, S. Ma, Q. Dai, J. Pendry, X. Zhang, and S. Zhang, Overcoming Losses in Superlenses with Synthetic Waves of Complex Frequency, 2023.

[66] F. Guan, Z. Lin, S. Chen, X. Wen, T. Li, and S. Zhang, High-order virtual gain for optical loss compensation in plasmonic metamaterials, Nat. Phys. (2026).

[67] T. Jiang, C. Zhang, R. Y. Zhang, Y. Yu, Z. Guan, Z. Wei, Z. Wang, X. Cheng, and C. T. Chan, Observation of non-Hermitian boundary induced hybrid skin-topological effect excited by synthetic complex frequencies, Nature Communications **15**, (2024).

[68] Y. Chen et al., Synthetic gain for electron-beam spectroscopy, Nat. Commun. (2026).

[69] F. Guan et al., Compensating losses in polariton propagation with synthesized complex frequency excitation, Nat. Mater. **23**, 506 (2024).

[70] K. Zeng, C. Wu, X. Guo, F. Guan, Y. Duan, L. L. Zhang, X. Yang, N. Liu, Q. Dai, and S. Zhang, Synthesized complex-frequency excitation for ultrasensitive molecular sensing, ELight **4**, 1 (2024).

[71] Z. Wu, S. Peng, Y. Wu, X. Liu, M. Li, X. Jin, G. Wang, S. Zhuang, and Q. Cheng, Synthesized complex-frequency excitation for enhanced terahertz time-domain spectroscopy sensitivity, Chinese Optics Letters **24**, 023001 (2026).

[72] A. Rogov and E. Narimanov, Space-Time Metamaterials, ACS Photonics **5**, 2868 (2018).
[73] T. Jiang, C. Zhang, R. Y. Zhang, Y. Yu, Z. Guan, Z. Wei, Z. Wang, X. Cheng, and C. T. Chan, Observation of non-Hermitian boundary induced hybrid skin-topological effect excited by synthetic complex frequencies, Nature Communications **15**, 1 (2024).
[74] Y. Chen et al., Synthetic gain for electron-beam spectroscopy, Nat. Commun. **17**, 1 (2026).
[75] F. J. Harris, On the Use of Windows for Harmonic Analysis with the Discrete Fourier Transform, PROCEEDINGS OF THE IEEE **66**, 51 (1978).
[76] S. M. Kay and S. L. Marple, Spectrum Analysis-A Modern Perspective, PROCEEDINGS OF THE IEEE **69**, 1380 (1981).
[77] A. H. Nuttall, Some Windows with Very Good Sidelobe Behavior, IEEE Trans. Acoust. **29**, 84 (1981).
[78] S. Zhang, D. A. Genov, Y. Wang, M. Liu, and X. Zhang, Plasmon-induced transparency in metamaterials, Phys. Rev. Lett. **101**, 047401 (2008).
[79] G. Zheng, H. Mühlenbernd, M. Kenney, G. Li, T. Zentgraf, and S. Zhang, Metasurface holograms reaching 80% efficiency, Nat. Nanotechnol. **10**, 308 (2015).